\documentclass[aps,prb,twocolumn,showpacs,superscriptaddress]{revtex4}
\usepackage{bm,amsmath,amssymb,latexsym,mathrsfs,enumerate}
\usepackage[mathcal]{euscript}
\usepackage{graphicx, subfigure}
\newcommand{\figwidth}{0.47\textwidth}

\begin{document}

\title{Solitons in isotropic antiferromagnets: beyond a sigma model.}

\author{ E. G. Galkina}
\affiliation{Institute of Magnetism, 03142 Kiev, Ukraine}
\affiliation{ Institute of Physics, 03028 Kiev, Ukraine}

\author{ A. Yu. Galkin}
\affiliation{ Institute of Metal Physics, 03142 Kiev, Ukraine}
\affiliation{Institute of Magnetism, 03142 Kiev, Ukraine}

\author{B. A. Ivanov}
\email{bivanov@i.com.ua} \affiliation{Institute of Magnetism, 03142
Kiev, Ukraine} \affiliation{National Taras Shevchenko University of
Kiev, 03127 Kiev, Ukraine}

\date\today

\begin{abstract}
Isotropic antiferromagnets shows a reach variety of magnetic
solitons with non-trivial static and dynamic properties.
One-dimensional  soliton elementary excitations have a periodic
dispersion law. For two-dimensional case, planar antiferromagnetic
vortices having non-singular macroscopic core with the saturated
magnetic moment are present. The dynamic properties of these planar
antiferromagnetic vortex are characterized by presence of a
gyroforce
\end{abstract}



\maketitle

\section{Introduction}

Magnetically ordered materials (magnets) are known as essentially
nonlinear systems and shows a large variety of localized nonlinear
excitations with finite energy, or solitons, see
Refs.~\onlinecite{Kosevich+All+R,Kosevich+All,MikStainer,
BarIvKhalat}.  It is sufficient to note kink-type solitons (domain
walls) which destroy long range order in one- dimensional systems;
magnetic vortices, which cause a  Berezinskii-Kosterletz-Thouless
transition in two-dimensional magnets with continuous
degeneration;\cite{Berezinsky72, Kosterlitz73} and also
two-dimensional localized solitons like Belavin-Polyakov
solitons,\cite{Belavin75} see for review
Refs.~\onlinecite{BarIvKhalat}. All these solitons were firstly
introduced in physics of magnets, and the development of soliton
concept for this particular region of physics is believed to be
important for modern nonlinear general physics of condensed matter
as well as for field models of high-energy physics, see
Ref.~\onlinecite{manton}.

To date, solitons in Heisenberg ferromagnets, whose dynamics are
described by the Landau--Lifshitz equation for the constant-length
magnetization vector, have been studied in details. From a
microscopic point of view description of such magnets is based on
the Landau-Lifshitz equation for a unit (normalized) magnetization
vector ${\rm {\bf m}}$, ${\rm {\bf m}}^2=1$, see
Refs.~\onlinecite{Kosevich+All+R,Kosevich+All,BarIvKhalat}.
Basically, for antiferromagnets one can use a set of two equations
for magnetizations of sublattices, which are unit vectors ${\rm {\bf
m}}_1 $ and ${\rm {\bf m}}_2 $, or, that is more convenient, their
irreducible combinations
\begin{equation}
\label{eq1}
{\rm {\bf m}}=({\rm {\bf m}}_1 +{\rm {\bf m}}_2 )/2,\;{\rm {\bf l}}=({\rm
{\bf m}}_1 -{\rm {\bf m}}_2 )/2,
\end{equation}
which are bound by constraint
\begin{equation}
\label{eq2}
({\rm {\bf m}},\;{\rm {\bf l}})=0,\;{\rm {\bf m}}^2+{\rm {\bf l}}^2=1.
\end{equation}
These variables naturally reflect the  symmetry inherent to
antiferromagnets, regarding sublattices rearrangement and they are
convenient for presentation of phenomenological energy of
antiferromagnet. However, the growing of the number of variables
essentially complicates the analysis, and within the framework of
this approach a few works have been done, we point out
Refs.~\onlinecite{Lelya,zvezin,EleKulAFM}.

A considerable progress in study  of non-linear dynamics of
antiferromagnets has been reached after obtaining of so-called
$\sigma -$model, which presents a dynamical equation for the
antiferromagnet vector ${\rm {\bf l}}$  see for review
Refs.~\onlinecite{Kosevich+All+R,Kosevich+All,MikStainer,
BarIvKhalat,IvKolFNT95,IvanovFNT05,barIvChUFN,bar-springer}. While
deducing the model one considers that ${\rm {\bf m}}$ is small,
${\rm {\bf m}}^2 \ll 1$, where ${\rm {\bf m}}$ is a slave variable
and is determined by the vector ${\rm {\bf l}}$ and its time
derivative $\partial {\rm {\bf l}}/\partial t$. $\sigma -$model
equations can be derived either directly from the Landau-Lifshitz
equations for sublattices magnetizations,\cite{IbarIv79,Mik80} or
phenomenologically, by account taken of symmetry
considerations.\cite{AndMarchUFN} It is a common belief that
description of nonlinear dynamics of antiferromagnets within $\sigma
-$model has the same level of universality as within the
Landau-Lifshitz equations for ferromagnets. At least it is
considered to be true for low-frequent dynamics in the longwave
approximation.

It is worth noting, the transition to $\sigma $-model is not
connected with any expansion over small amplitudes of deviations of
the vector ${\rm {\bf l}}$ from the equilibrium position. Hence,
$\sigma -$model is highly nonlinear. Since within $\sigma -$model
${\rm {\bf l}}$ is considered as a unit vector, this model is a
typical nonlinear chiral model, in which a non-linearity is
determined a geometric condition ${\rm {\bf l}}^2=1$. However, it
turns out that an isotropic $\sigma -$model as a nonlinear system is
to a certain extent quite ``poor''. In particular, for a
non-localized nonlinear wave of a structure $l_x +il_y =l_0 \cdot
\exp ({\rm {\bf kr}}-\omega t)$, $l_z =\sqrt {1-l_0^2 }
=\mbox{const, }$where the wave amplitude $l_0 <1$ can be not small,
the frequency $\omega $ for a given ``wave vector'' ${\rm {\bf k}}$
is independent on the wave amplitude $l_0 $. As well, in this system
there are no traveling-wave solitons, which are most indicative
nonlinear excitations. Note that for the case of anisotropic
antiferromagnets with a uniaxial or rhombic magnetic anisotropy such
traveling-wave solitons are present, they describe moving domain
walls, see Refs.~\onlinecite{barIvChUFN,bar-springer}. It is
interesting to sort whether the abovementioned absence of two
specific nonlinear effects is an intrinsic property of an isotropic
antiferromagnet or it appeared due to approximations done during
transition to $\sigma -$model.

To answer this question  it is necessary to proceed from a full
system of equations for two vector variables ${\rm {\bf m}}$ and
${\rm {\bf l}}$, bound by the relation (\ref{eq2}). Such an
analysis, in principle, is considerably complicated as one has to
deal with four nonlinear equations, rather than two angular
variables for the unit vector ${\rm {\bf l}}$ as for $\sigma
-$model. However, we can limit our consideration to analysis of some
concrete class of solutions in order to confirm the presence of
solitons.

In this article,  a class of solutions in a simple model of an
antiferromagnet with consideration of only isotropic exchange
interaction is pointed out. In such a solution the vector ${\rm {\bf
m}}$ is parallel to some direction and change its length only, while
the vector ${\rm {\bf l}}$ turns around it within some plane. It is
appropriate to call these solutions as ``planar''. Within the class
of such solutions, consistent description of properties of nonlinear
waves and soliton dynamics is done.

The article is organized as following.  In the Section 2 a model is
formulated and effective equations of spin dynamics in terms of
${\rm {\bf m}}$ and ${\rm {\bf l}}$ without application of typical
for $\sigma -$model approximations are presented, and the integrals
motion are obtained. On the basis of these equations in the Section
3 the soliton structure is calculated for a one-dimensional case.
The analysis of the dispersion law of solitons done in the Section 4
demonstrates these one-dimensional stable solitons are magnetic
analogies of Lieb states known from one-dimensional Bose gas
model.\cite{Lieb} Further in the Section 5 two-dimensional solitons
describing magnetic vortices are analyzed.

\section{Model, effective equations of spin dynamics and
conservation laws.}

Dynamical equations for the vectors ${\rm {\bf m}}$ and ${\rm {\bf
l}}$ can be written as follows
\begin{multline}
\label{eq3}
 \hbar S\frac{\partial {\rm {\bf m}}}{\partial t}=\left( {{\rm {\bf
m}}\times \frac{\delta W}{\delta {\rm {\bf m}}}} \right)+\left( {{\rm {\bf
l}}\times \frac{\delta W}{\delta {\rm {\bf l}}}} \right), \\
 \hbar S\frac{\partial {\rm {\bf l}}}{\partial t}=\left( {{\rm {\bf
l}}\times \frac{\delta W}{\delta {\rm {\bf m}}}} \right)+\left(
{{\rm {\bf m}}\times \frac{\delta W}{\delta {\rm {\bf l}}}} \right),
\end{multline}
where $W=W[{\rm {\bf m}},{\rm {\bf l}}]=\int {w\{{\rm {\bf m}},{\rm
{\bf l}}\}(d^dx/a^d)} $ is the energy functional of an
antiferromagnet, which is presented here for a magnet with a
hypercubic lattice with dimension $d$, $w=w\{{\rm {\bf m}},{\rm {\bf
l}}\}$  is the energy density, which depends on the vectors ${\rm
{\bf m}}$ and ${\rm {\bf l}}$ and their spatial derivatives. In the
standard expansion on gradients with account taken of Eq.
(\ref{eq2}), a general expression for $w$ in the case of a purely
isotropic antiferromagnet takes the form
\begin{equation}
\label{eq4} w=JS^2{\rm {\bf m}}^2+\frac{1}{2}A_1 a^2S^2\left(
{\nabla {\rm {\bf m}}} \right)^2+\frac{1}{2}A_2 a^2S^2\left( {\nabla
{\rm {\bf l}}} \right)^2\; ,
\end{equation}
where $J$ is the effective homogeneous exchange constant,  the
parameters $A_1 $ and $A_2 $ are determined by exchange integrals
within one sublattice and between sublattices, respectively, $S$ is
an atomic spin and  $a$ is the lattice constant. For this energy,
the magnetization vector ${\rm {\bf m}}$ equals to zero in the
ground state. It is worth noting, for the model (\ref{eq4}) with the
$A_1 =0$ the $\sigma -$model representation is exact, while the
values of both constants are important for the soliton solutions for
antiferromagnet. Below, we will not specify relations between the
constants $A_1 $ and  $A_2 $ and their connections with some
microscopic spin model.

The equations (\ref{eq3})  have the obvious integral of motion, the
whole system energy $E$, values of which coincide with the value of
$W[{\rm {\bf m}},{\rm {\bf l}}]$ calculated for some concrete
solution, and the field momentum ${\rm {\bf P}}$, which would be
described below. For an isotropic problem the total spin value ${\rm
{\bf S}}^{(tot)}=\int {S{\rm {\bf m}}d^dx/a^d} $ is also an integral
of motion. It can be derived from a dynamical equation for the spin
density ${\rm {\bf m}}$, with usage of energy form (\ref{eq4}), that
gives
\begin{equation}
\label{eq5}
\hbar \frac{\partial {\rm {\bf m}}}{\partial t}=-\mbox{div}\left[ {S\left(
{{\rm {\bf m}}\times A_1 \nabla {\rm {\bf m}}} \right)+S\left( {{\rm {\bf
l}}\times A_2 \nabla {\rm {\bf l}}} \right)} \right].
\end{equation}

This expression determines the conservation law of the total spin
${\rm {\bf S}}^{(tot)}$ in differential form. Its analysis allows
also to point out a concrete exact class of solutions for the full
set of equations (\ref{eq5}). Let the vector ${\rm {\bf m}}$ and its
time and space derivatives at the initial moment of time are
parallel to some direction, which can be chosen as the z axis. The
equation (\ref{eq2}) demonstrate that in this case the vector ${\rm
{\bf l}}$ and its derivatives lie in the perpendicular (x, y) plane.
In virtue of (\ref{eq5}) such geometry remains for subsequent
moments of time, i.e. dynamical equations for an antiferromagnet
allow a planar solution in the form of ${\rm {\bf m}}\vert \vert
{\rm {\bf e}}_z ,\;\,{\rm {\bf l}}\bot {\rm {\bf e}}_z $. Accounting
the constraint (\ref{eq2}) the vectors ${\rm {\bf m}}$ and ${\rm
{\bf l}}$ can be parameterized by two angular variables,
\begin{equation}
\label{eq6}
{\rm {\bf m}}={\rm {\bf e}}_z \sin \mu ,\;\,{\rm {\bf l}}=\cos \mu \left(
{{\rm {\bf e}}_x \cos \varphi +{\rm {\bf e}}_y \sin \varphi } \right),
\end{equation}
where ${\rm {\bf e}}_x $ and ${\rm {\bf e}}_y $ are unit vectors
directed along x and y axis, respectively. The initial isotropy of
the problem in this case manifest itself in arbitrary directions of
axis ${\rm {\bf e}}_z $, ${\rm {\bf e}}_x $ and ${\rm {\bf e}}_y $,
specific for the planar solution.

An important characteristic of the planar solution is that the system
dynamics with new variables$\mu ,\;\varphi $ can be described by a simple
Lagrangian
\begin{equation}
\label{eq7} L=\int {\frac{d^dx}{a^d}\left( {-\hbar S\frac{\partial
\varphi }{\partial t}\sin \mu \,-w} \right)} \; ,
\end{equation}
where $w$ is the energy density \eqref{eq4} presented through
angular variables,
\begin{multline}
\label{eq8} w=JS^2\sin ^2\mu +\frac{1}{2}A_2 a^2S^2\cos ^2\mu \left(
{\nabla \varphi }
\right)^2+ \\
 +\frac{1}{2}a^2S^2\left[ {A_1 \cos ^2\mu +A_2 \sin ^2\mu } \right]\left(
{\nabla \mu } \right)^2\;.
\end{multline}
Lagrangian approach allows one to obtain  an expression for linear
momentum of the magnetic excitation ${\rm {\bf P}}$, which is a
total field momentum of corresponding field,
\begin{equation}
\label{eq9} {\rm {\bf P}}=\hbar S\int {\frac{d^dx}{a^d}(\nabla
\varphi )\sin \mu } \; .
\end{equation}
The dynamical part of the Lagrangian (\ref{eq7}) and the expression
for momentum (\ref{eq9}) contain singularities connected with non
differentiability of the azimuthal angle $\varphi $. This property
of the variable $\varphi $ plays a significant role in description
of vortices dynamics in ferromagnets.\cite{PapanikoVihri} In our
case, the presence of this singularity will also manifest itself
essentially in description of solitons dynamics, either
one-dimensional or two-dimensional, see Sections 4,5.

\section{ Non-linear waves and one-dimensional solitons. }

Lets us consider dynamics  of a simple magnetization wave
propagating along some direction, say, the x-axis, with the velocity
$v$. For such wave, $\mu =\mu (\xi )$, $\varphi =\varphi (\xi )$,
$\xi =x-vt$. For an analysis of such solutions it is easier to start
with the spin conservation equations (\ref{eq5}), which can be
integrated once and then gives an apparent relation of $\varphi
'=d\varphi /d\xi $ (in this Section, the derivative over $\xi $ is
denoted by prime) and $\mu $ in the following form
\begin{equation}
\label{eq10}
\varphi '=\frac{\hbar v\sin \mu +C_1 }{a^2SA_2 \cos ^2\mu }
\end{equation}
where $C_1 $ is an arbitrary constant. Using this expression it is
possible to introduce the Lagrange equation $\delta L/\delta \mu$ in
the form of the second order ordinary differential equation for $\mu
\left( \xi \right)$. It is easy to demonstrate that this equation
has the first integral, and for $\mu \left( \xi \right)$ one can
obtain a simple equation with separating variables. Hence the
problem allows a general analysis of nonlinear waves depending on
one parameter, the wave velocity $v$, and containing, in a general
case, two arbitrary constants $C_1 $ and $C_2 $. The explicit
solution of this equation can be presented in elliptic functions.

First of all we are interested in soliton solutions for which, far
from a soliton, at $\xi \to \pm \infty $, $\mu \left( \xi \right)$
turns zero, while $\varphi \left( \xi \right)$ has constant value.
Therefore we consider only the case $C_1 =0$. Then the equation for
$\mu \left( \xi \right)$ in soliton solution acquires the following
form
\begin{multline}
\label{eq11} a^2\left[ {A_1 \cos ^2\mu +A_2 \sin ^2\mu }
\right]\left( {\mu '}
\right)^2= \\
 =\sin ^2\mu \left( {2J-\frac{\hbar ^2v^2}{a^2S^2A_2 \cos ^2\mu }} \right)\;.
\end{multline}

Let us discuss properties of such soliton solutions. A simple
analysis demonstrates that the soliton velocity has an upper limit,
the value $c=2\sqrt {JA_2 } \hbar S/a$, which coincides with phase
velocity of linear excitations (magnons) for antiferromagnet. This
is a rather natural condition for traveling-wave solitons. It is
worth noting that $c$ does not depend on constant $A_1 $, thus, it
can be obtained in the framework of $\sigma -$model.  However
soliton states exist only at $A_1 >0$. The latter is a formal
confirmation of the fact that for their analysis one should go
beyond this model.

The soliton solution of this equation can only be written through
elliptic functions.  The structure of the planar solitons in
antiferromagnets, as well as the energy dependence on the soliton
velocity, is quite common to that for solitons in spin nematic
state.\cite{IvKhymynJETP07} Hence, we will not discuss it in details
and limit ourselves with its qualitative analysis. First of all, the
form of the solution depends highly on the value of soliton velocity
$v$. If the velocity $v$ is nearly $c$, the soliton amplitude $\mu
_{\max } $ is small, proportional to $\sqrt {c-v} $. The maximal
value of $\mu _{\max } $ is reached at the zero soliton velocity.

As follows from the equations (\ref{eq10}) and (\ref{eq11}), the
values of $\varphi $ at the right and left of the soliton differ by
a certain value $\Delta \varphi $. In the case $A_1 =A_2 $ the value
of $\Delta \varphi =\pi $ and it is independent of the soliton
velocity. For any other relation between $A_1 $ and $A_2 $, this
limit value $\Delta \varphi =\pi $ appears at zero soliton velocity,
but $\Delta \varphi <\pi $ for $v\ne 0$ and it vanishes at $v\to c$.
In principle, all these features are common to that for a so-called
rotary waves for easy plane ferromagnets,  see for review
Refs.~\onlinecite{Kosevich+All+R,Kosevich+All}, or the so-called
dark solitons, which are well known in nonlinear
optics.\cite{KivsharDark}

The energy of a soliton is one of most  important soliton
characteristics. Using Eqs. (\ref{eq10}) and (\ref{eq11}), the
energy density $w$ (\ref{eq8}) can be easily present through the
function $\mu (\xi )$ only. It is convenient to write down the
soliton energy $E$ as a definite integral over $\mu $ from $\mu =0$
till the maximal value $\mu _{\max } $. Again, the explicit value of
this integral can be written through a simple but long combination
of elliptic integrals only. The exception is the limit case $A_1
=A_2 $, for which the explicit form for soliton energy as a function
of its velocity can be written as a simple square root dependence,
\begin{equation}
\label{eq12}
E=E_0 \sqrt {1-\frac{v^2}{c^2}}
\end{equation}
where $c$ is the spin wave speed, $E_0 =2aS^2\sqrt {2JA} $ is the maximal
soliton energy, corresponding to the zero soliton velocity $v=0$, in the
case $A=A_1 =A_2 $.

\section{Semiclassical quantization of one-dimensional solitons. }

The soliton energy $E$ and momentum $P$ are the for are the most
natural soliton characteristics and the dependence $E(P)$ is the
basis for their semiclassical
quantization.~\cite{Kosevich+All+R,Kosevich+All} Within this
approach, $E(P)$ dependence can be considered as a dispersion n law
for quantum nonlinear elementary excitations that are described by
solitons. Usually, this dependence, which is found from classical
solutions, well reflects the properties of the corresponding quantum
results.

As has been noted above, the energy is maximal for a stationary
soliton with $v=0$, and it vanishes at $v\to c$. The concrete
dependence can be easily found by numerical estimates of
corresponding integral, see Ref.~\onlinecite{IvKhymynJETP07}.
Concerning soliton momentum, the situation is not so easy. It is
worth noting, the equation (\ref{eq10}) gives $d\varphi /d\xi =0$ at
$v=0$ and $C_1=0$, that formally means zero value of momentum. On
the other hand, for any $v\ne 0$ the soliton momentum $P(v)$ is
finite, and the limit value of the function $P(v)$ at $v\to 0$ is
also finite. For example, for simplest case $A_1 =A_2 $ one can
easily find $P=(\hbar S/a)\cdot \arccos \left( {v/c} \right)$, that
gives $P\to \pm \pi \hbar S/2a$ at $v\to \pm 0$. Combining this
dependence with Eq. (\ref{eq12}), one can present the dispersion
relation for this particular case as a periodic function,
\begin{equation}
\label{eq13} E=E_0 \cdot \left| {\sin \left( {\frac{\pi P}{2P_0 }}
\right)} \right|,\;P_0 =\frac{2\pi \hbar S}{a}\; .
\end{equation}
with universal period $P_0 $. The question appears, whether or not
these features, the periodicity of the dispersion relation and the
value of period are model independent.

In principle, this problem can be overcame by detail investigation
of the behavior of the soliton solution at small velocities, see
Refs.~\onlinecite{Kosevich+All+R,Kosevich+All} for more details. On
the other hand, it is useful to present a general model-free
discussion, as it has been done for domain walls in
ferromagnets.\cite{Galkina,IvMik} Let discuss this problem in more
details; moreover, it will be useful for the description of
dynamical properties of vortex-like two-dimensional solitons.

Indeed, according to Eq. (\ref{eq9}), the soliton momentum contains
a singularity related to the presence of the gradient of the
azimuthal angle $\varphi$. Such singularity is an internal property
of the Lagrangian, see Eq. (\ref{eq7}). It becomes clear if we
parameterized the spin variables of the planar solution through a
three-dimensional vector ${\rm {\bf R}}$, ${\rm {\bf
R}}=(X,\;Y,\;Z)=\left( {m,\;l_x ,\;l_y } \right)$, whose components
represent nontrivial variables for the planar solution, namely, a
magnetization $m=m_z $ and two non-zero projections of the vector
${\rm {\bf l}}$. Then the density of the dynamical part of the
Lagrangian (\ref{eq7}) can be written as
\begin{equation}
\label{eq14} {\rm {\bf A}}({\rm {\bf R}})\frac{\partial {\rm {\bf
R}}}{\partial t},\;{\rm {\bf A}}({\rm {\bf R}})=\frac{\hbar
S}{a}\cdot \frac{Z(Y{\rm {\bf e}}_x -X{\rm {\bf e}}_y
)}{R(X^2+Y^2)},\;R=\vert {\rm {\bf R}}\vert
\end{equation}
where the vector $\mathbf{A}$ has a singularity along the $Z-$axis.
This Lagrangian coincides with that for a charged particle with the
coordinate $\mathbf{R}$ in a magnetic field with the vector
potential $\mathbf{A }$. This representation also holds true for a
ferromagnet in terms of the Landau--Lifshitz equation; however,
expressions for $\mathbf{A }$ in these two cases are different. We
can readily show that, although the expressions for ${\rm {\bf
A}}={\rm {\bf A}}({\rm {\bf R}})$ are different for the cases of an
antiferromagnetic planar solution and a ferromagnet, for Eq.
(\ref{eq14}) we have ${\rm {\bf B}}=\mbox{rot}{\rm {\bf A}}=\hbar S
{\rm {\bf R}}/aR^3$ Thus, as in the case of a ferromagnet, Eq.
(\ref{eq14}) describes the vector potential of a magnetic monopole
located at the origin. Therefore, the expressions for a momentum $P$
of one-dimensional soliton can be obtained by the substitution
$\partial {\rm {\bf R}} / \partial t \to -v\partial {\rm {\bf R}}
\partial  \xi $; it can be
reduced to the same form as for a soliton in a ferromagnet by gauge
transformation. We then can use the same method as in Refs.
\onlinecite{Galkina,IvMik}.

The formula for the one-dimensional soliton momentum $\mbox{
}P\mbox{ }=\mbox{ }\int {{\rm {\bf A}}({\rm {\bf R}})d{\rm {\bf R}}}
$, contains a singularity and is not invariant with respect to the
gauge transformations of the vector potential $\mathbf{A}$. However,
it is important that the vector $\mathbf{B}$ does not contain
singularities on the sphere ${\rm {\bf R}}^2=1$. Whence, it follows
that the difference in the momenta of two different soliton states
is a gauge-invariant quantity. Indeed, every soliton (e.g., solitons
with different velocities) can be associated with a trajectory
connecting certain points ${\rm {\bf R}}^{(-)}$ and ${\rm {\bf
R}}^{(+)}$ lying in the equator of the sphere ${\rm {\bf R}}^2=1$
(circle $Z=0$ or $m =0)$. In this case, the momentum of this soliton
is specified by the integral $\int {{\rm {\bf A}}d{\rm {\bf R}}} $
over this trajectory going from the point ${\rm {\bf R}}^{(-)}$ to
the point ${\rm {\bf R}}^{(+)}$. Although different solitons (e.g.,
solitons with different velocities) have different values of the
variable $\varphi  \; {\rm g}$at infinity, all of them have $m=0$ at
infinity; that is, they finish at the equator of the sphere ${\rm
{\bf R}}^2=1$. In this line, the integrand is exactly zero;
therefore, the ends of the illustrating trajectories of two solitons
that finish at different points in the great circle can be connected
by a segment lying in this circle and can be considered to be
closed. It is clear that the difference in the momenta of the two
solitons is determined by the integral over the closed contour
$\oint {{\rm {\bf A}}d{\rm {\bf R}}\mbox{ }} $ bound by the
trajectories describing these solitons. According to the Stokes
theorem, this integral can be written as a flux of a vector ${\rm
{\bf B}}=\mbox{rot}{\rm {\bf A}}$ through the surface enclosed by
this contour. Therefore, the difference in the momenta of two
soliton states $\Delta P$ can be represented in the gauge invariant
form
\begin{equation}
\label{eq15} \mbox{ }\Delta P\mbox{ }=\frac{\hbar S}{a}\int {{\rm
{\bf B}}d{\rm {\bf S}}} \mbox{= }\frac{\hbar S}{a}\int {\cos \mu
\;d\mu \; d\varphi }
\end{equation}
Here the variables $\pi/2-\mu $ and $\varphi $ can be considered as
the standard spherical coordinates for the vector ${\rm {\bf R}}$,
and the integral is taken over the region on the sphere bound by the
trajectories corresponding to these two solitons. It is natural to
choose the equator as the line corresponding to $P=0$, to which the
soliton trajectories tend asymmetrically as the soliton amplitude
decreases; this corresponds to $E\to 0$ and $v\to c$. The maximum
soliton energy corresponds to a trajectory that passes through the
``north pole'' of the sphere; for this pole, we have $P=P_0 /2$ and
$E=E_{\max } $. The $V(P)$ and $E(P)$ dependencies are then
qualitatively restored. Indeed, all trajectories corresponding to a
soliton velocity in the range from $v=c$ to $v=0$ or to a soliton
momentum from zero to $P_0 /2$ fill the gap between these two limit
trajectories. Hence, the momentum increases continuously when going
from the trajectory near the equator and when approaching the
limiting trajectory with $\pm P_0 /2$. As a soliton trajectory moves
further in the second half of the upper hemisphere, the energy
decreases and the momentum increases until this trajectory reaches
the equator. Here, the energy is $E = 0$, the momentum (with
allowance for the choice of its reference point) is determined by
integral (\ref{eq15}) over the entire upper hemisphere, and $P=P_0
$.

Thus, as for domain walls in a ferromagnet,\cite{Galkina,IvMik} a
true periodic $E\left( P \right)$ dependence appears for a planar
solitons in an antiferromagnets due to the topological properties of
the Lagrangian. This fact should lead to specific features in forced
soliton motion, e.g., to oscillating soliton motion under the action
of a constant force (Bloch oscillations) as was discussed in details
by Kosevich in Ref.~\onlinecite{KosevichBloch}.

\section{Two-dimensional solitons - antiferromagnetic vortices
with ferromagnetic core }

Let us consider the static and dynamic properties of two-dimensional
topological solitons on the basis of the model given by Eq.
(\ref{eq4}). For two-dimensional planar solitons the Lagrange
equation for the variable $\varphi$ takes the form
\begin{equation}
\label{eq16}
\hbar S\frac{\partial \mu }{\partial t}\sin \mu =A_2 a^2\nabla [\sin ^2\mu
\left( {\nabla \varphi } \right)],
\end{equation}

In the static case, according to this equation, a two-dimensional
solution can be taken in the form
\begin{equation}
\label{eq17}
\varphi =m\chi +\varphi _0 ,\;\mu =\mu (r),
\end{equation}
where $r$ and $\chi$ are the polar coordinates in the plane of the
system and $\varphi _0 $ is an arbitrary angle. To have a continuous
distribution of the vectors ${\rm {\bf m}}$ and ${\rm {\bf l}}$, the
number $m$ should be integer. The structure of the vortex core is
determined by the function $\mu (r)$ for which the ordinary
differential equation can be obtained
\begin{multline}
\label{eq18} \left[ {1+\kappa ^2\sin ^2\mu } \right]\cdot \left(
{\frac{d^2\mu
}{dr^2}+\frac{1}{r}\frac{d\mu }{dr}} \right)- \\
 -\sin \mu \cos \mu \left[ {\frac{1}{l_0^2 }-\frac{A_2 m^2}{A_1 r^2}-\kappa
^2\cdot \left( {\frac{d\mu }{dr}} \right)^2} \right]=0,
\end{multline}
$\kappa ^2=\left( {A_2 -A_1 } \right)/A_1 , \; l_0 =a\sqrt {A_1 /2J}
$ is the characteristic length scale. If the condition $A_1 =A_2 =A$
holds, Eq. (\ref{eq18}) by substitution $\mu \to \pi /2-\theta $
transforms into the equation describing the vortex in easy plane
ferromagnet, see Refs.~\onlinecite{Kosevich+All+R,Kosevich+All}. It
is easy to show that at $r \gg l_0 $ the quantity $\mu $ reaches its
equilibrium value $\mu =0$, and the behavior near the coordinate
origin is a power law: $\mu (r)-\pi /2\propto r^m$. Such power
dependence is characteristic of a out-of-plane vortex in
ferromagnets. Thus, at the center of the planar antiferromagnetic
vortex a nonsingular saturated core with approximately ferromagnetic
order is formed, and in the vortex center the magnetization takes
its maximal value, see Fig.1.

It is easy to show that the energy  of a planar antiferromagnetic
vortex, as well as of other topological defects, has a weak
(logarithmic) divergence with an increase in the system size $L$, it
can be written as
\begin{equation}
\label{eq19} E=m^2\frac{\pi A_2 S^2 a^2}{2}\cdot \ln \left(
{\frac{L}{\eta l_0 }} \right)\; ,
\end{equation}
where $\eta $ is a numerical factor on the order of unity. Hence,
the vortex with $m=\pm 1$ has the minimal energy, and further we
will discuss only this case.

\begin{figure}[!t]
\includegraphics[width=\figwidth]{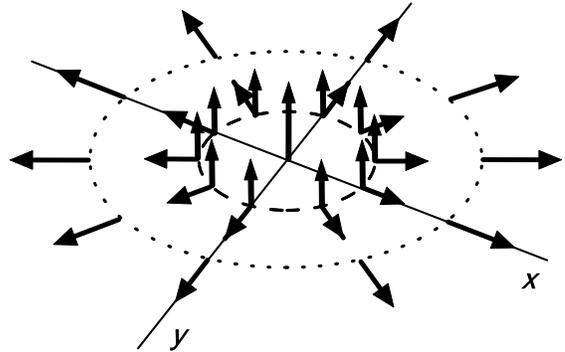}
\caption{\label{fig1} Schematic distribution of the vector ${\rm
{\bf l}}$ (in-plane arrows with wide heads) and the vector ${\rm
{\bf m}}$ (vertical arrows) in the planar antiferromagnetic vortex
with the vorticity $m=1$. The core border, chosen as the line with
$\mu =\pi /4$, is marked by the dashed line circle. The outermost
circle (formally, the circle with $r\to \infty $, with the value of
$\mu =0)$ is schematically shown by the dotted line circle.}
\end{figure}

It is interesting to compare the energy of this planar
antiferromagnetic vortex with that for vortices in easy-plane
antiferromagnets. In principle, planar antiferromagnetic vortices
contains a ferromagnetic core with almost parallel sublattice
magnetizations ${\rm {\bf m}}_1 $ and ${\rm {\bf m}}_2 $. On the
first glance, this costs too much energy comparing with that for
easy-plane antiferromagnetic vortices. But this energy difference
enters the logarithmic multiplier, see Eq. (\ref{eq19}). Thus, this
difference is unimportant for many physical applications; for
example, the only logarithmic dependence of the energy on the system
size is manifesting the temperature of the
Berezinskii-Kosterletz-Thouless transition in two-dimensional
systems. Thus, both kinds of vortices can be important for a
description of such transitions for real antiferromagnets.

Let us describe dynamic properties of the planar antiferromagnetic
vortex, which are also nontrivial. In the framework of the $\sigma
-$model, the solution describing any soliton freely moving with a
velocity of $v<c$ can be obtained from the known immobile solution
by the Lorentz transformation with the chosen speed $c$. However,
the $\sigma -$model is inapplicable for the planar antiferromagnetic
vortex considered above. Analysis shows that the motion of the
planar antiferromagnetic vortex is possible only against the
background of ``spin flux,'' i.e., a nonzero value of $\nabla
\varphi ={\rm {\bf k}}$ at infinity. Vortex velocity ${\rm {\bf v}}$
and ${\rm {\bf k}}$ are related as $\hbar S{\rm {\bf v}}=2a^2A_2
\cdot {\rm {\bf k}}\cdot $; this relation can be derived using the
same method as in Ref.~\onlinecite{NikSonin} for a vortex in a
ferromagnet. On the other words, far from the core of moving vortex
the ``condensate'' is non-uniform, with $\nabla \varphi ={\rm {\bf
k}}\propto {\rm {\bf v}}=d{\rm {\bf X}}/dt$. Thus, the total energy
of the system containing a freely moving planar antiferromagnetic
vortex diverges as ${\rm {\bf v}}^2L^2$, $L^2$ is a system area, and
the notion of the local inertial mass losses meaning. This property
is known for vortices in ferromagnets or superfluid systems and
corresponds to freezing of vortices in the condensate, see for
review  Ref.~\onlinecite{BarIvKhalat,bar-springer}.

The problem of the forced motion of the  planar antiferromagnetic
vortex can be considered by analyzing the field momentum ${\rm {\bf
P}}$. Similar to a ferromagnet, Eq. (\ref{eq8}) includes the
non-differentiable expression, which leads to nontrivial features of
the momentum of the topological soliton in these
systems.\cite{PapanikoVihri} It is most simple to use the method
proposed in Ref.~\onlinecite{IvanovSteph89} and to calculate the
quantity $d{\rm {\bf P}}/dt$ in the leading approximation in the
vortex velocity ${\rm {\bf v}}$. To this end, it is sufficient to
use the immobile solution given by Eq. (\ref{eq10}) with a change of
$\mathbf{r}$ by $\tilde {\mathbf{r}}$, where ${\rm {\bf \tilde
{r}}}={\rm {\bf r}}-{\rm {\bf X}}(t)$, ${\rm {\bf X}}={\rm {\bf
X}}(t)=X{\rm {\bf e}}_x +Y{\rm {\bf e}}_y $ is a coordinate of the
vortex center. In this approximation, $\mu =\mu (\tilde {r})$,
$\varphi =m\tilde {\chi }$,  $\tilde {r}=\vert {\rm {\bf \tilde
{r}}}\vert $ and $\tilde {\chi }=\arctan [(y-Y)/(x-X)]$. Having in
mind some general features of the vortex motion for the models with
gyroscopic dynamics like in Lagrangian of Eq. (\ref{eq7}), let us
start with the general form of these term as in Eq. (\ref{eq14}),
not using the concrete form of the vector-potential ${\rm {\bf A}}$.

In the leading approximation on  the vortex velocity ${\rm {\bf
v}}$, the $\alpha -$th component of the time derivative of the
vortex momentum, $d{\rm {\bf P}}_0 /dt$ with the taken into account
the conditions $\partial {\rm {\bf R}}/\partial t=-v_\alpha
(\partial {\rm {\bf R}}/\partial x_\alpha )$ can be rewritten as
\begin{equation}
\label{eq20}
\frac{dP_{0,\alpha } }{dt}=\int {d^2x\cdot } \frac{\partial R_i }{\partial
x_\alpha }\frac{\partial R_j }{\partial x_\beta }v_\beta \left(
{\frac{\partial A_j }{\partial R_i }-\frac{\partial A_i }{\partial R_j }}
\right)
\end{equation}

As for the momentum of one-dimensional soliton, this expression
contains gauge-invariant quantity ${\rm {\bf B}}=\mbox{rot}{\rm {\bf
A}}$, $\partial A_j /\partial R_i -\partial A_i /\partial R_j
=\varepsilon _{ijk} (\mbox{rot}{\rm {\bf A}})_k $, instead of
vector-potential ${\rm {\bf A}}$ as itself. Then the direct
calculation yields, $d{\rm {\bf P}}_0 /dt=G\cdot ({\rm {\bf e}}_z
\times {\rm {\bf V}})$. Here the gyroconstant $G$, as well as the
linear momentum for one-dimensional solitons (\ref{eq15}), can be
presented in the gauge invariant form $\mbox{ }G\mbox{ }=\hbar S\int
{{\rm {\bf B}}d{\rm {\bf S}}} $, as a flux of the vector ${\rm {\bf
B}}$ through the area of the sphere ${\rm {\bf R}}^2=1$,
corresponding to the vortex, that gives $G=2\pi \hbar S/a^2$.

\section{Conclusion.}
Thus, beyond the $\sigma -$model approximation the isotropic
antiferromagnets shows a reach variety of magnetic solitons with
non-trivial static and especially dynamic properties. For
one-dimensional magnet, soliton elementary excitations with a
periodic dispersion law exists. These soliton excitations have
common features with the so-called Lieb states,\cite{Lieb} which are
well known in many condensed matter models. For two-dimensional
case, planar antiferromagnetic vortices having non-singular
macroscopic core with the saturated magnetic moment are found. The
dynamic properties of these planar antiferromagnetic vortex are also
unusual. Moving planar antiferromagnetic vortex is subjected to the
gyroscopic force $G\cdot [{\rm {\bf e}}_z ,{\rm {\bf V}}]$,
equivalent to the Lorentz force for a charged particle in the
uniform magnetic field, it is well known for vortices in easy-plane
ferromagnets and superfluid systems, and is observed in experiments
on the motion of magnetic bubbles and Bloch lines.\cite{MalozSlon}
In contrast, gyroforce never appears in Lorentz-invariant $\sigma
-$model equation; for a usual vortex in an antiferromagnet the
gyroscopic force can be induced only by the strong external magnetic
field and is absent for $H $= 0.\cite{IvShekaAFM} It is worth
noting, both these non-trivial dynamical characteristics, period in
dispersion law $P_0 $ and gyroconstant $G$, can be written through
gauge-invariant expressions of the common form. These quantities are
independent on exchange integrals and depends only on a spin value
$S$ a single crystal parameter, namely, the interatomic distance
$a$.

We thank V.G. Bar'yakhtar and A.S. Kovalev for useful discussions of the
results. This work was supported in part by the INTAS Foundation, project
INTAS-05-1000008-8112 and by the joint grant Ô25.2/081 from Ministry of
Education and Science of Ukraine and Ukrainian State Foundation of
Fundamental Research.

\end{document}